\documentclass[conference, 10pt]{IEEEtran}
\usepackage{tikz}
\usepackage{textcomp}
\usepackage{hyperref}
\usepackage{lipsum}

\newcommand\copyrighttext{%
  \footnotesize This work has been presented at the 2019 IEEE International Symposium on Antennas and Propagation and USNC-URSI Radio Science Meeting, July 2019. \\ \textcopyright 2019 IEEE. Personal use of this material is permitted.
  Permission from IEEE must be obtained for all other uses, in any current or future
  media, including reprinting/republishing this material for advertising or promotional
  purposes, creating new collective works, for resale or redistribution to servers or
  lists, or reuse of any copyrighted component of this work in other works.
}
\newcommand\copyrightnotice{%
\begin{tikzpicture}[remember picture,overlay]
\node[anchor=south,yshift=10pt] at (current page.south) {\fbox{\parbox{\dimexpr\textwidth-\fboxsep-\fboxrule\relax}{\copyrighttext}}};
\end{tikzpicture}%
}

\ifCLASSINFOpdf
\else
\fi
\usepackage[cmex10]{amsmath}
\usepackage{amsfonts}
\usepackage{multirow}
\makeatletter
\newcommand{\xRightarrow}[2][]{\ext@arrow 0359\Rightarrowfill@{#1}{#2}}
\makeatother
\usepackage{algorithm}
\usepackage{algorithmic}

\newcommand{\beq}{\begin{equation}}
\newcommand{\eeq}{\end{equation}}

\usepackage{graphicx}
\usepackage{color}
\usepackage{placeins}
\usepackage{float}
\usepackage{hyperref}
\usepackage{tabularx,colortbl}

\begin{document}
%
\title{Parallel Direct Domain Decomposition Methods (D$^3$M) for Finite Elements}

\author{\IEEEauthorblockN{Javad Moshfegh, Dimitrios G. Makris, and Marinos N. Vouvakis}
\IEEEauthorblockA{Department of Electrical and Computer Engineering, University of Massachusetts, Amherst, MA, USA}}


%


\maketitle
\copyrightnotice

\begin{abstract}
A parallel direct solution approach based on domain decomposition method (DDM) and directed acyclic graph (DAG) scheduling is outlined. Computations are represented as a sequence of small tasks that operate on domains of DDM or dense matrix blocks of a reduced matrix. These tasks can be statically scheduled for parallel execution using their DAG dependencies and weights that depend on estimates of computation and communication costs. Performance comparison with MUMPS 5.1.2 on electrically large problems suggest up to 20\% better parallel efficiency, 30\% less memory and  slightly faster in run-time,  while maintaining the same accuracy.
\end{abstract}


%
\IEEEpeerreviewmaketitle

\section{Introduction}

Direct solution methods (factorization-based) and high performance parallel computing are growing trends in electromagnetic computations. When it comes to  discretization of frequency-domain Maxwell's equations via Finite Element Methods (FEM), the resulting linear systems are  sparse and indefinite thus very hard to solve with direct factorization methods, and even more challenging to efficiently parallelize. State-of-the-art sparse matrix direct solvers such as MUMPS [\ref{ref:amestoy2001}] and PARDISO [\ref{ref:schenk2004solving}] although very reliable and fast at least on smaller scale problems, they tend to not scale favorably, and have low parallel efficiency and very high memory footprint. 

Recent advances in parallel dense direct solvers, have shifted toward parallel implementation that rely on Directed Acyclic Graph (DAG) scheduling approaches [\ref{ref:buttari2009class}] that can achieve highly efficient asynchronous parallel execution. However, adaptation of such schemes to sparse matrices is very hard or even  impractical. 

Direct Domain Decomposition Method (D$^3$M), introduced in [\ref{ref:moshfegh2016direct}--\ref{ref:vouvakis2018sparse}], offer a reliable memory efficient sparse direct solver for FEM that could deliver better parallel performance. In D$^3$M, a sparse FEM matrix is reduced via an ``embarrassingly parallel" step into a block-wise sparse matrix. A special block LDL$^T$, well suitable for  block directed acyclic graph (B-DAG) task scheduling asynchronous parallel execution, is used to solve the reduced system, before another step of ``embarrassingly parallel"  primal unknown recovery. Using this parallelization method, one can achieve significant parallel scaling improvement and time saving.

\section{Parallel D$^3$M}

\begin{figure}[thpb]
\begin{center}
\includegraphics[width=3.5in]{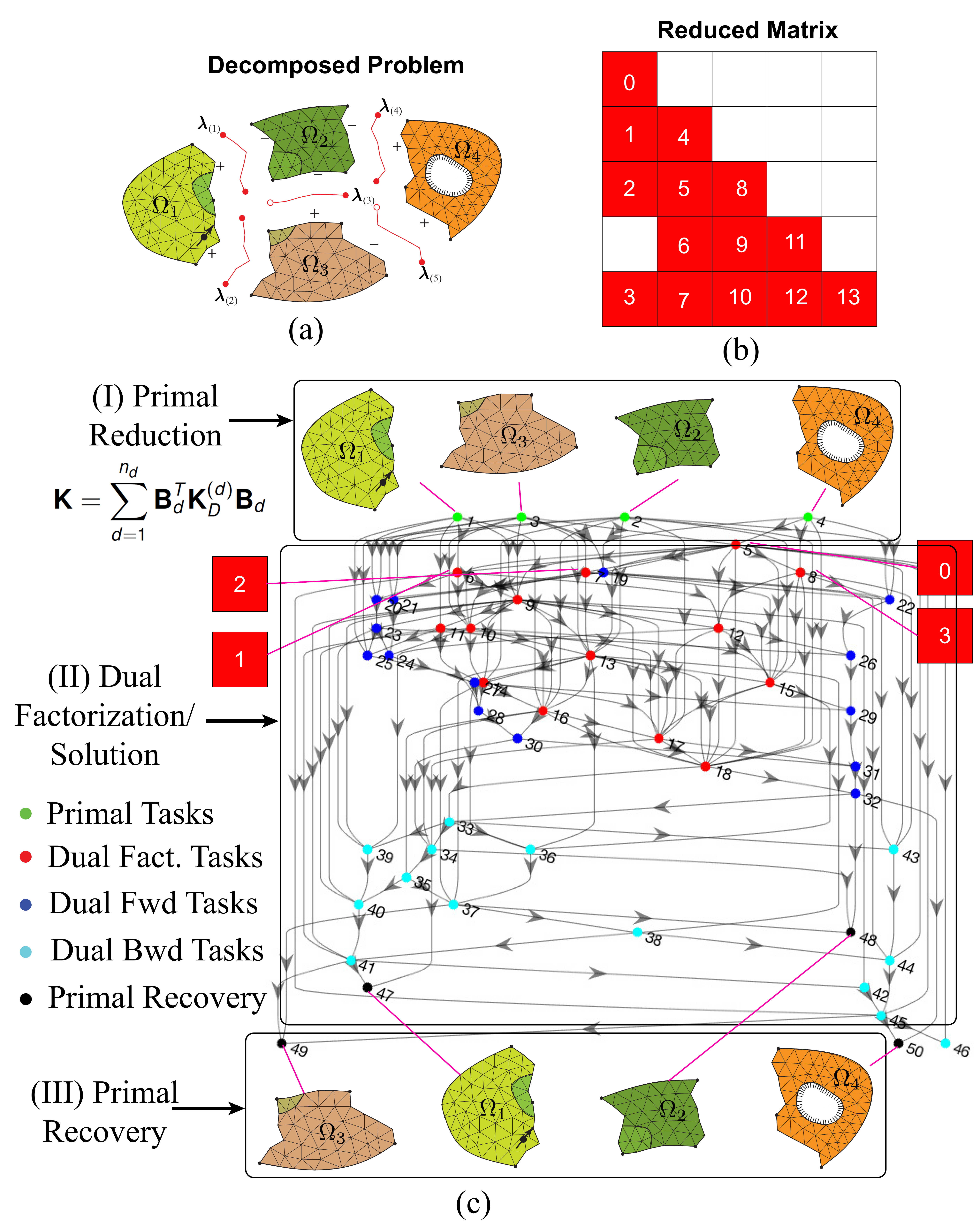}
\vspace{-17pt}
\caption{Overview of the proposed parallel methodology. (a) Cross-section of a sample four-domain problem; (b) The sparsity of the reduced matrix ${\bf K}$; (c) The task graph.}\label{fig:approach}
\end{center}
\end{figure}

The D$^3$M algorithm steps is decomposed into set of instructions called primitive tasks (ptask), e.g. `dense update' or `dense triangular solve'. To increase data locality, ptasks are agglomerated into tasks that share local memory.  A dependence analysis of those tasks must be performed to determine the spatio/temporal distribution of tasks into cores. This is done by symbolic simulation of the sequential D$^3$M algorithm to generate a B-DAG, called task graph.
For example, the task graph of a four-domain problem is shown in Fig. \ref{fig:approach}.

\vspace{-10pt}

D$^3$M task graph has three parts, the primal reduction, dual factorization and solution, and primal recovery part. In primal reduction a domain $\Omega_i$ is represented by a task responsible for sparse factorization of the regularized  FEM matrix, the inversion on its interfaces (DtN map computation), and generation of dual domain matrices. In the dual factorization and solution step each non-zero block of the dual matrix is represented by a task responsible for possible dense update, dense triangular solver, and dense factorization. Finally in the primal recovery part each domain $\Omega_i$ is again represented by a `task' responsible for recovery of the primal unknowns and possible computation of scattering matrix S or far-field.

The weights of the B-DAG graph associated with the  primal tasks are estimated using K-nearest neighbor method, whereas the weights for dual tasks are estimated using benchmarked level 3 BLAS operations: GEMM, sytrf, and triSolve.

The weighted B-DAG is the input of a list scheduling heuristic algorithm, which maps the tasks statically to the available processor units and determine their execution order. Asynchronous communication aided by manual progression  \cite{manualp} used to ensure the overlap between  communication and computation. Finally, the parallel code executes the assigned tasks on each processor.

\section{Numerical Results}
\begin{figure}[htpb]
\begin{center}
\includegraphics[width=3.5in]{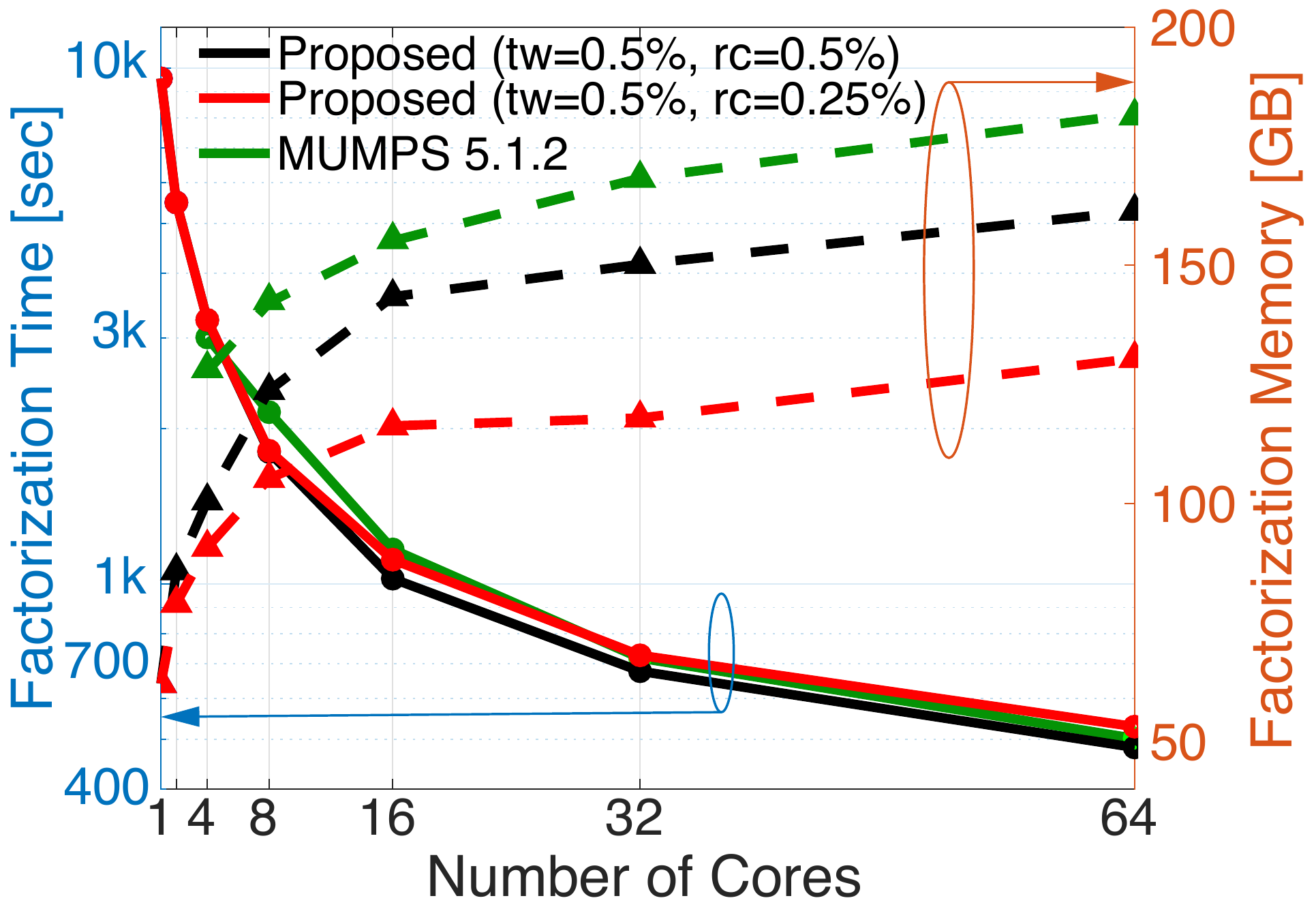}
\vspace{-10pt}
\caption{Factorization time and memory for a 7$\lambda$ sphere problem.}\label{fig:time_memory}
\end{center}
\end{figure}

\begin{figure}[htpb]
\begin{center}
\includegraphics[width=3.in]{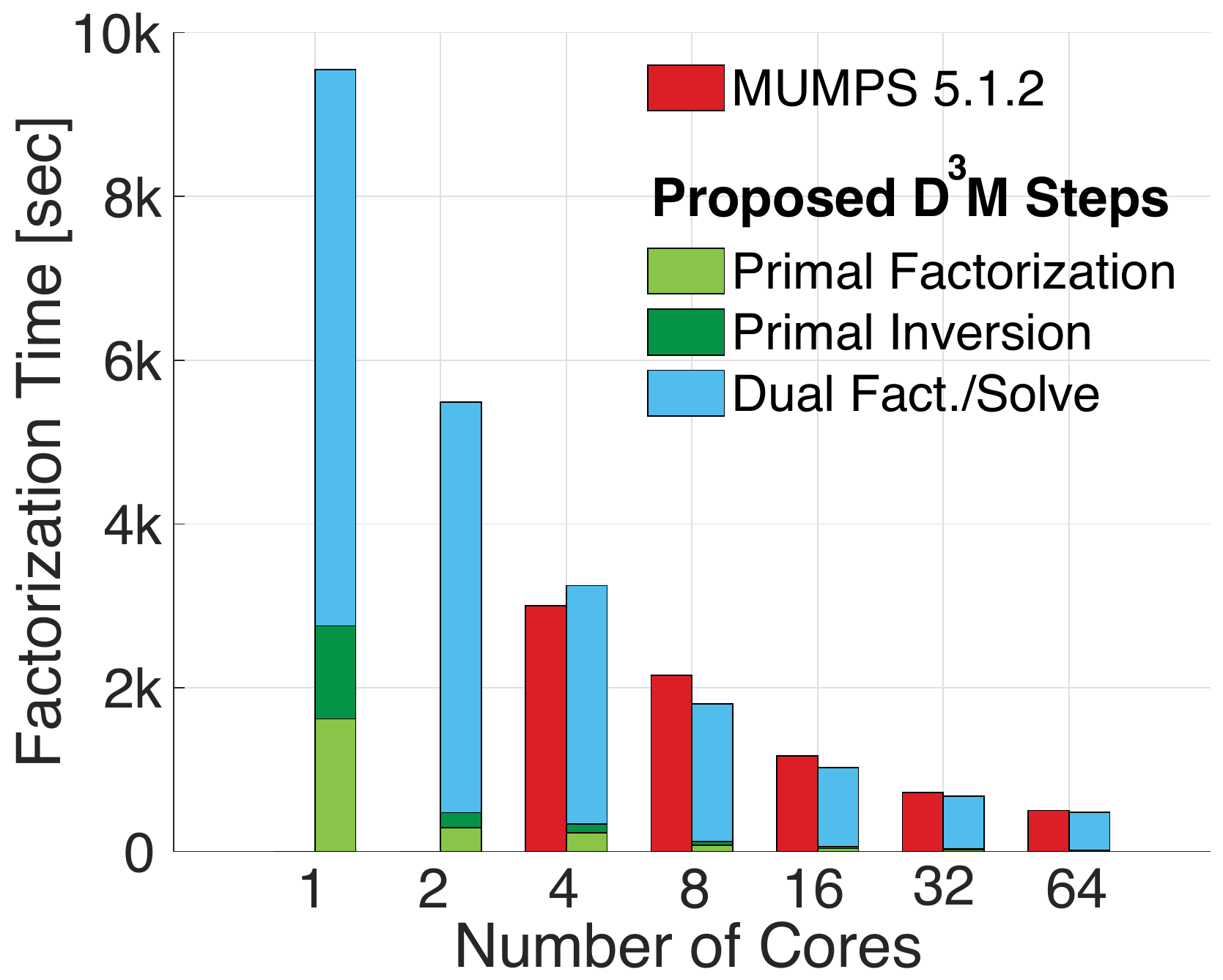}
\vspace{-10pt}
\caption{Time break-down in proposed D$^3$M for a 7$\lambda$ sphere problem.}\label{fig:bar}
\end{center}
\end{figure}

\begin{figure}[htpb]
\begin{center}
\includegraphics[width=3.in]{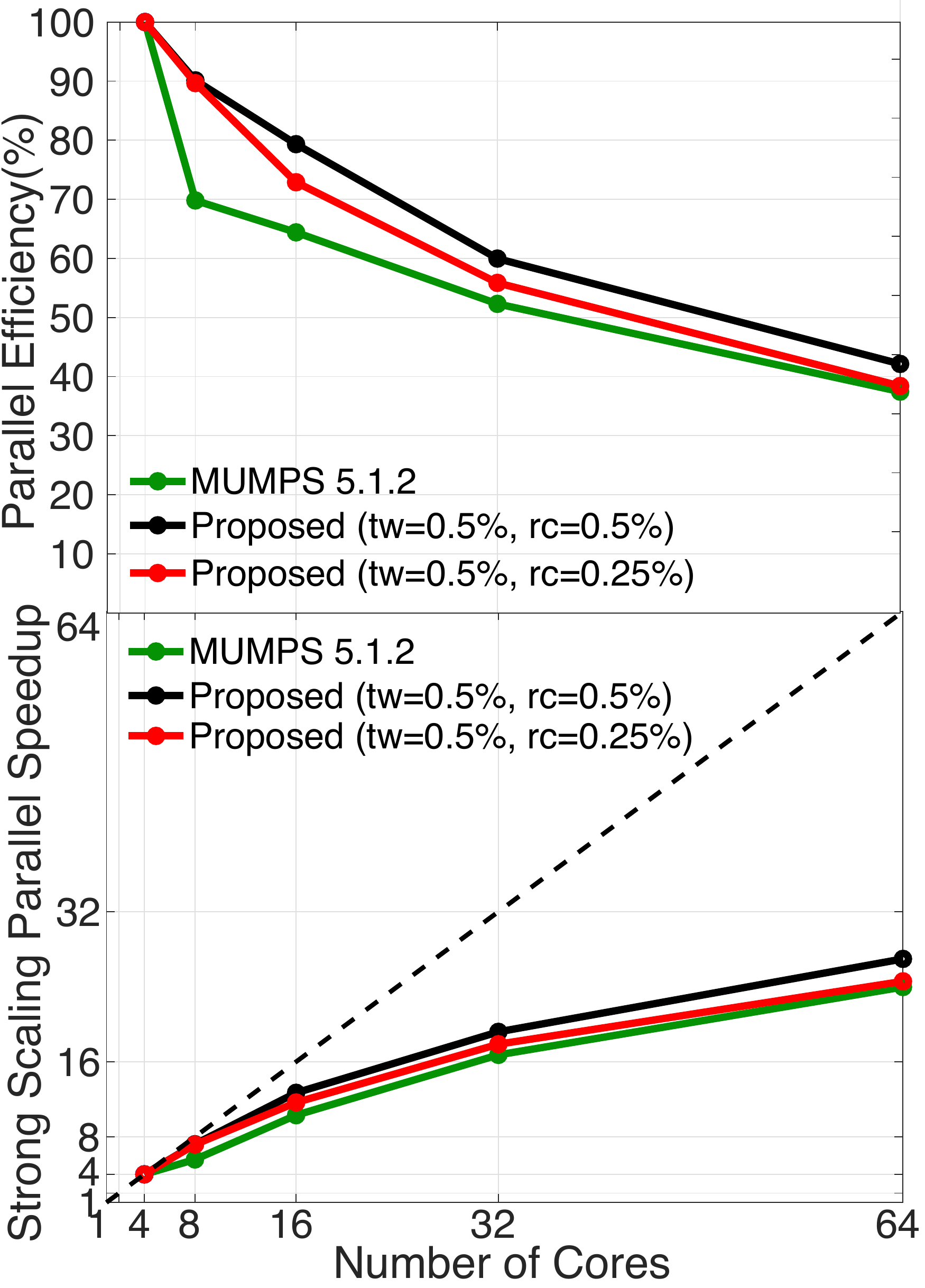}
\vspace{-10pt}
\caption{Strong parallel scaling for 7$\lambda$ sphere problem.}\label{fig:total}
\end{center}
\end{figure}

The scattering of a dielectric sphere ($\epsilon_r$=4) with diameter 7$\lambda_0$, partitioned into 1000 domains, is considered. The problem is discretized in 534,516 tetrahedra and 3,411,490 second order Nedelec tangential vector FEM unknowns. A series of runs with  increasing number of distributed cores is performed to compare the strong scalability of the proposed method to MUMPS 5.1.2 with METIS 5.1.0 reordering. Factorization time and memory of the two methods are compared in Fig. \ref{fig:time_memory}, where green lines represent the MUMPS results.  The proposed method has some tuning parameters that mainly control memory. The tuning set-up in black curve offers the fastest time, and about 10$\%$ less memory than MUMPS 5.1.2,  while the red curve offers better memory, approximately 30\% less memory than MUMPS 5.1.2, but at almost same speed.  The time breakdown for the main steps of D$^3$M method are plotted in Fig. \ref{fig:bar} highlighting that the dual factorization/solution step dominates. Strong scaling parallel speedup and parallel efficiency are compared in Fig. \ref{fig:total}. The proposed D$^3$M has up to 20\% better parallel efficiency than MUMPS 5.1.2.

\end{document}